\documentstyle[prl,twocolumn,aps]{revtex}
\begin{document}
\input{epsf.tex} \epsfverbosetrue

\draft

\twocolumn[\hsize\textwidth\columnwidth\hsize\csname @twocolumnfalse\endcsname

\title{Dipole-Mode Vector Solitons}

\author{Juan J. Garc\'{\i}a-Ripoll$^{1}$, V\'{\i}ctor M.
  P\'erez-Garc\'{\i}a$^{1}$, Elena A. Ostrovskaya$^{2}$, and Yuri S.
  Kivshar$^{2}$}

\address{$^{1}$Departamento de Matem\'aticas, Escuela T\'ecnica Superior de
  Ingenieros Industriales, \\
  Universidad de Castilla-La Mancha 13071 Ciudad Real, Spain}

\address{$^{2}$Optical Sciences Centre, The Australian National University,
  Canberra ACT 0200, Australia}

\date{\today}

\maketitle

\begin{abstract}
  We find a new type of optical vector soliton that originates from trapping of
  a dipole mode by a soliton-induced waveguide.  These solitons, which
  appear as a consequence of the vector nature of the two component system, are
  more stable than the previously found optical vortex-mode solitons and represent a
  new type of extremely robust nonlinear vector structure.
\end{abstract}

]

\narrowtext


Perhaps one of the most desirable goals of Optics is the development of purely
optical devices in which light can be used to guide and manipulate light
itself. This motivation explains the growing interest in self-guided beams (or
{\em spatial optical solitons}) and the recent theoretical and experimental
study of spatial solitons and their interactions \cite{stegeman}. It is not only the reconfigurable and steerable soliton-induced waveguides created in a bulk
medium that are of particular practical interest, but also a spatial soliton that guides another beam (of a different polarization or frequency). This may become a completely new object, a {\em vector soliton}, with an internal structure and new dynamical properties which yield surprising results for the stability of such an object even in simplest cases.

Complex phenomena induced by the {\em vector nature} of nonlinear wave
equations arise in many fields of Physics and are already firmly placed in the
realm of condensed matter physics, the dynamics of biomolecules, and nonlinear optics.  Recently an interest in these structures and the theoretical
possibilities they offer has been renewed because of their experimental
realization in different physical contexts. For instance, vector phenomena have
been observed in Bose-Einstein condensation, with vortices in multicomponent
condensates \cite{vortices} or in nontrivial topological defects due to
interspecies interaction \cite{monopolos}. Finally, vectorially active
optical media are also being currently investigated because of many new
characteristics they provide, as compared to scalar systems \cite{colet}.

In this Letter we study a vector object formed by two optical beams that
interact incoherently in a bulk nonlinear medium. If the nonlinearity of the
medium is self-focusing, an isolated beam, under proper conditions, will form a
self-trapped state - {\em a spatial optical soliton} \cite{stegeman}. Such a soliton changes the refractive index of the medium and creates a stationary
effective waveguide. A second beam of a much lower intensity is subjected to
the induced change of the refractive index and can be trapped as a localized
mode of that waveguide.

From linear optical waveguide theory, we expect that a radially symmetric
waveguide can host different types of modes with more elaborate geometries
[Figs. 1(a-c)].  However, at higher intensities of the trapped beam one must
regard the two beams as components of a {\em vector soliton}, self-trapped by a
self-consistent change of the refractive index induced by {\em both beams}. In
this case we cannot treat the shapes of the beams as independent and it is not
trivial to conclude whether we may obtain states which are a composition of a
lowest-order state in one component and a high-order state in the other one.
\begin{figure}
  \setlength{\epsfxsize}{8cm} \centerline{ \epsfbox{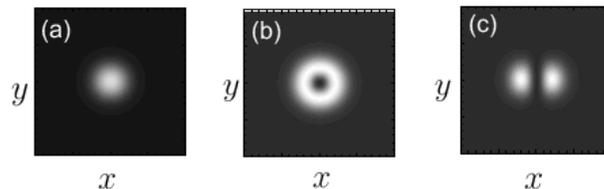}}
  \caption{
    Intensity distribution of (a) fundamental, (b) vortex, and (c) dipole
    modes supported by a radially symmetric fundamental soliton-induced waveguide.}
  \label{fig1}
\end{figure}
Recalling a previous work on the existence and stability of two-component one-dimensional solitons \cite{prl}, one may expect to find at least
two types of such complex objects in the two-dimensional case.  The first
type is the recently discussed two-dimensional vector soliton
\cite{vortex} which has a node-less shape [e.g., as shown in Fig. 1(a)] in
the first component and a vortex on the second one [Fig. 1(b)]. The second
 type introduced here maintains a similar shape for the first
component, while the second beam develops a node along a certain direction
[Fig.  1(c)] forming what we call a {\em dipole-mode vector soliton}.

The purpose of this Letter is two-fold. First, we discuss the stability of the
vortex-mode vector solitons and show that these objects are {\em linearly unstable} and can decay into dipole-mode vector solitons. Secondly, we prove that these
dipole-mode solitons exist for an ample range of relative intensities of their
components and show that they survive both small and large amplitude
perturbations, their propagation dynamics resembling that of two spiraling beams \cite{spiral}.

We would like to emphasize here that {\em both results are highly nontrivial}.
While it is commonly believed that asymmetric solitary waves possess a higher
energy, and should be {\em a priori} unstable, our results demonstrate that
{\em the opposite is true}: An excited state with an elaborate geometry may
indeed be more stable than a radially symmetric one and, as such, would be a
better candidate for experimental realization. We stress here
 that the recently discovered method of creating multi-component spatial optical solitons in a
photorefractive medium \cite{christ} would allow a simple and direct
verification of our theory, including the questions of soliton existence and
stability.

The outline of the Letter is as follows. First we formulate the model for our
system. Next, we proceed with the study of
vortex-mode vector solitons in the reasonable parameter range of the model. We recall previous studies on the issue of existence of such states \cite{vortex,md} and add new results arising from a linear stability analysis and numerical simulations of the dynamics of
these solitons. Having concluded that the vortex-mode vector solitons are
 linearly unstable, we proceed with the study of the dipole variant.  We obtain
a continuous family of stationary solutions in the same parameter range. In this case, our tools for the linear stability analysis give no conclusive results, but
numerical simulations of highly perturbed states show a periodic, stable
evolution. Finally we conclude and summarize the implications of this work.

{\em The model}.- We consider two incoherently interacting beams propagating
along the direction $z$ in a bulk, weakly nonlinear {\em saturable} optical
medium. The model corresponds, in the isotropic approximation, to the
experimentally realized solitons in photorefractive materials. The problem is
described by the normalized, coupled equations for the slowly varying beam
envelopes, $E_1$ and $E_2$. The normalized dynamical equations for the
envelopes of two incoherently interacting beams can, in this case, be
approximately written in the form \cite{spiral,christ}:
\begin{equation}
\label{nls}
i \frac{\partial E_{1,2}}{\partial z} + \Delta_{\perp} E_{1,2} + \frac{ E_{1,2}}{ 1+ |E_{1,2}|^2 + |E_{2,1}|^2} = 0,
\end{equation}
where $\Delta_{\perp}$ is the transverse Laplacian. Stationary solutions of
Eqs. (\ref{nls}) can be found in the form
$E_1=\sqrt{\beta_1}\,u(x,y)\,\exp({i\beta_1z})$,
$E_2=\sqrt{\beta_1}\,v(x,y)\,\exp({i\beta_2z})$, where $\beta_1$ and $\beta_2$
are two independent propagation constants. Measuring the transverse coordinates
in the units of $\sqrt{\beta_1}$, and introducing the ratio of the propagation
constants, $\lambda = (1-\beta_2)/(1-\beta_1)$, from Eqs. (\ref{nls}) we derive
a system of stationary equations for the normalized envelopes $u$ and $v$:
\begin{eqnarray}
\label{sat nls}
\Delta_{\perp} u - u + uf(I)= 0, \\ \nonumber
\Delta_{\perp} v - \lambda v +vf(I) = 0, 
\end{eqnarray}
where $f(I)=I(1+s I)^{-1}$, $I=u^2+v^2$, and $s=1-\beta_1$ plays the
role of a saturation parameter. For $s=0$, this system describes the Kerr
nonlinearity. In this paper we will work with intermediate values of
saturation, around $s=0.5$.

{\em Vortex-mode solitons}.- First, we look for radially symmetric solutions
$u(x,y)=u(r)$, $w(x,y)=w(r)\exp(im \phi)$, in which the second component
carries a topological charge, $m$, and we assume that the $u$ component has no
charge.  In this case, Eqs. (\ref{sat nls}) take the form:
\begin{eqnarray}
\label{rad}
\Delta_{\rm r} u - u + uf(I)= 0, \\ \nonumber
\Delta_{\rm r} v - (m^2/r^2)v -\lambda v +vf(I) = 0, 
\end{eqnarray}
where $\Delta_{\rm r}=(1/r)(d/dr)(r d/dr)$. The fundamental, bell-shaped
solutions with $m=0$ only exist at $\lambda=1$. In the remaining region of the
parameter plane $(s,\lambda)$, the solutions carrying a topological charge
($m=\pm 1$) in the second component exist. Solutions of this type for the
saturable nonlinearity are found and thoroughly investigated in \cite{md}, and
in \cite{vortex} - for the so-called threshold nonlinearity. The families of
these {\em radially symmetric}, two-component vector solitons are characterized by a
single parameter $\lambda$, and at any fixed value of $s$, the border of their
existence domain is determined by a cut-off value, $\lambda_c$. A two-component
trapped state exists only for $\lambda > \lambda_c$.  Near the cutoff point
this bell-shaped state can be presented as a waveguide created by the
$u-$component guiding a small-amplitude mode $v$. Away from the cut-off,
the amplitude of the $v-$component grows, and the resulting vector soliton
develops a {\em ring-like shape}. An example of the ring-shape vortex-mode for our model is
presented in Figs. 2(b). An important physical characteristic of vector
solitons of this type is the total power defined as $P = P_u + P_v = 2 \pi
\int^\infty_0\,(u^2 + v^2)rdr$, where the partial powers $P_u$ and $P_v$ are
the integrals of motion for the model (\ref{nls}). The dependencies
$P(\lambda)$, $P_v(\lambda)$, and $P_u(\lambda)$ for a fixed $s$ completely
characterize the family of vector solitons [a typical example is shown in Fig.
2(a)] for $\lambda > \lambda _c$.
\begin{figure}
  \setlength{\epsfxsize}{8cm} \centerline{ \epsfbox{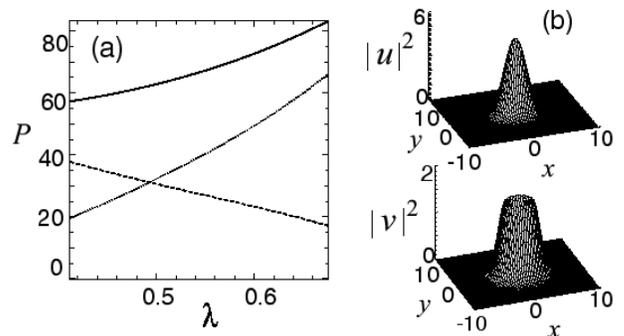}}
  \caption{
    (a) Power diagram for the vortex-mode vector soliton family for $s=$0.5 (solid --
    total power, dashed -- $P_u$, dotted -- $P_v$) and (b) typical intensity
    profiles of the soliton components at $\lambda=0.6$.}
  \label{fig2}
\end{figure}
{\em Stability analysis.- }Similar to the well studied case of the (1+1)-D
vector solitons (see, e.g., Ref. \cite{prl}), the vortex-mode soliton is 
associated, in the linear limit, with a soliton-induced waveguide supporting a
higher-order mode.  It is therefore tempting to draw the analogy between the
higher-order (1+1)-D two-hump solitons and (2+1)-D ring-shape solitons. Given
the established stability of the multi-hump one-dimensional structure in a {\em
  saturable medium} \cite{prl}, this line of thought would lead us
(erroneously!) to conclude that the vortex-mode vector solitons in our model
should be linearly stable.

To show that the above conclusion is wrong, we have performed a linear
stability analysis of the two-dimensional vortex-mode vector solitons. Our technique
consists in linearizing Eqs. (\ref{nls}) around the vortex solution and
evolving them with completely random initial conditions \cite{Soto}. Usually,
the solution will be a linear combination of modes evolving with some real
frequencies, $\mu$. However, if the linear equation bears modes with complex
eigenvalues [$\mu=\mathrm{Re}(\mu)+\mathrm{i} \mathrm {Im}(\mu)$], we expect an exponential growth of our
random data, with convergence to the invariant space of one of these eigenvalues. This method allows us to extract the most unstable eigenvalue and
its associated manifold in a similar way to the classical analysis of
Lyapunov exponents in unstable systems.

Our linear stability analysis has proved that, although saturation does have a
strong stabilizing effect on the ring vector solitons \cite{md}, all vector
solitons of this type {\em are linearly unstable}. In Fig. 3(a), where we plot a typical
dependence of the eigenvalue $\mu$ of the most unstable mode on the soliton
parameter $\lambda$ for a fixed $s$, we see that the growth rate of the
instability tends to zero at the cut-off point of the vortex mode [cf. Fig.
2(a) and 3(a)].  This behaviour is consistent with the inherent stability of the fundamental scalar soliton in a saturable medium. Elsewhere the growth rate is positive and increases when the increasing intensity of the vortex mode.
\begin{figure}
  \setlength{\epsfxsize}{8cm} \centerline{ \epsfbox{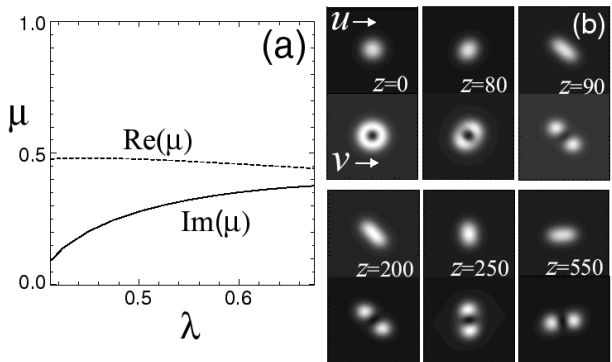}}
  \caption{
    (a) Eigenvalue of the leading unstable mode for vortex-mode solution
    ($s=0.5$) and (b) typical evolution of the vortex-mode soliton near cut-off
    ($s=0.65$, $\lambda=0.6$). Shown are intensity distributions of the $u$-
    and $v$- components in the $(x,y)$ plane [$-10< (x,y) <10$].}
  \label{fig3}
\end{figure}
In Figs. 3(a-b) we compare the linear stability analysis with dynamical
simulations of the vortex-mode soliton near cutoff, perturbed with random
noise. The instability, although largely suppressed by saturation, triggers the
decay of the soliton into a {\em dipole structure} [as shown in Fig. 3(b)] for
even a small contribution of the charged mode. The dipole demonstrates
astonishing persistence for large propagation distances as a rotating and
radiating {\em pulsar state}.
\begin{figure}
  \setlength{\epsfxsize}{8cm} \centerline{ \epsfbox{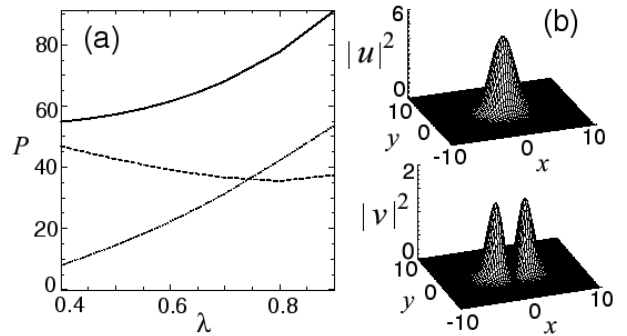}}
  \caption{
    (a) Power diagram of the dipole-mode vector soliton family for $s=0.5$ (solid --
    total power, dashed -- $P_u$, dotted -- $P_v$), and (b)
    typical intensity profiles of the soliton components at 
     $\lambda=0.6$.}
  \label{fig4}
\end{figure}

{\em Dipole-mode solitons.- }It is apparent from the above analysis that the
{\em dipole-mode} vector soliton should be more stable than the vortex-mode
soliton. We have identified the existence domain of these solitons by solving
numerically the stationary equations (\ref{sat nls}) to find localized asymmetric solutions carrying a dipole mode [as shown in Fig. 1(c)] in the $v-$component. One characteristic example of the dipole-mode soliton family is shown in Fig. 4(a) for a fixed $s$. Our linear stability analysis for these solutions does not converge to a particular
value of $\mathrm{Im}(\mu)$. This indicates that either the eigenvalues of the
unstable modes, if they exist, are extremely small, or that the unstable modes
have shapes which are only too weakly excited by the random perturbations.

To obtain further information on the dynamical stability of dipole-mode vector
solitons, we have propagated numerically different perturbed dipole-mode
solitons for distances up to several hundreds of $z$ units, or diffraction
lengths. To put these numbers into physical perspective, we note that, in
current experiments on solitons in photorefractive materials, the typical
crystal length is $\approx 20$ $mm$, whereas $z=100$ in our model corresponds to the soliton propagation length $\approx 20-40$ $mm$.

We have performed two types of numerical experiments. First, we have found that
small perturbations or random noise lead to bounded oscillations of the vector
soliton, which retains its shape. This shows that the dipole-mode vector
solitons should be stable enough for experimental observation. On the other
hand, strong perturbations, such as a large disproportion of the humps of the
dipole or a relative displacement of the components, may alter the shape of the
dipole. Such a perturbation, in the process of evolution, is typically
transfered from one component to the other in a robust, periodic way. It is
clear from our simulations [Fig. 5] that in this case the dynamics is
similar to that of two beams spiraling together but with initial zero angular momentum
\cite{spiral}.
\begin{figure}
  \setlength{\epsfxsize}{7.5cm} \centerline{ \epsfbox{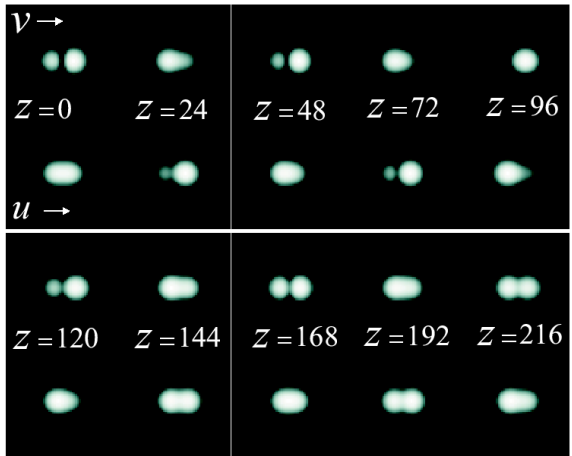}}
  \caption{
    Evolution of a dipole vector soliton, $(\lambda,s)$ $=$ $(0.7,0.5)$, after
    moving 25$\%$ of the total power from one hump to another. Shown are
    intensity distributions for the $u-$ and $v-$ components in the $(x,y)$ plane [$-7<(x,y)<7$].}
  \label{fig5}
\end{figure}
From the evidence above it is clear that the unstable modes of a dipole
 vector soliton, if they exist, should be rare and hard to excite. Calculations
of the linear stability spectrum around the exact soliton solution should
provide a complete answer to the stability question, which will be the subject
of our further research.

Summing up the stability results, we can state that the dipole-mode vector
solitons are extremely robust objects. According to our numerical tests, they
have a typical lifetime of several hundred diffraction lengths and survive a
wide range of perturbations. This is true even for vector states with a large
power in the dipole component ($P_w > P_u$), i.e. in the parameter region which
cannot be safely reached for vortex-mode solitons, since the latter become
unstable much sooner.

Finally, it is important to mention that there exist other physical models,
where {\em stable, radially asymmetric, dipole solitary waves} play an
important role in the nonlinear wave dynamics. All these models, however,
possess {\em scalar}, or one-component, structures. The most famous examples
are the Larichev-Reznik soliton, a localized solution of the Charney equation
for Rossby waves \cite{reznik}, and the dipole Alfv\'en vortex soliton in
an inhomogeneous plasma \cite{petvi}. Another type of dipole solitary waves is
found as a two-soliton bound state in a non-local \cite{gorshkov} or
anisotropic nonlocal \cite{mamaev} media due to anomalous interaction of two
solitons with opposite phase.  Nevertheless, the physics behind all known
dipole solitary waves and corresponding single-component nonlinear models {\em
  differ drastically} from the problem considered above.  Therefore, the
dipole-mode vector soliton we describe in this Letter is a genuinely new type
of solitary waves in a {\em homogeneous isotropic} bulk medium, a phenomenon that may
occur in many other physical applications.

In conclusion, we have analyzed the existence and stability of radially
symmetric and asymmetric higher-order vector optical solitons in a saturable
nonlinear bulk medium, and predicted a new type of optical soliton associated with the dipole mode guided by a soliton-induced waveguide. We have
demonstrated that solitons carrying a topological charge are linearly unstable and,
as a consequence, they may decay into dipole-mode solitons. There is also
strong evidence of the stability of these dipole-mode solitons. We believe that
all the effects predicted in this Letter, including the existence and stability
of solitons, can be easily verified in experiments with
photorefractive media.

J.J. G-R. thanks Optical Sciences Center, ANU for warm hospitality during his
stay in Australia. V.M.P-G.  and J.J.G-R. are partially supported by DGICYT
under grant PB96-0534.

\end{document}